\documentclass{PoS}

\usepackage[numbers]{natbib}
\usepackage{lineno}

\usepackage{graphicx}

\usepackage{siunitx}
\newcommand{\eV}{\electronvolt}
\newcommand{\keV}{\kilo\electronvolt}
\newcommand{\MeV}{\mega\electronvolt}
\newcommand{\GeV}{\giga\electronvolt}
\usepackage[version=4]{mhchem}

\newcommand{\compair}{\emph{ComPair}}
\newcommand{\fermiLAT}{\emph{Fermi}-LAT}
\newcommand{\amego}{\emph{AMEGO}}

\newcommand{\eg}{e.g.}

\title{Subsystem Development for the All-Sky Medium Energy Gamma-ray Observatory (AMEGO) prototype}

\ShortTitle{Subsystem Development for the AMEGO prototype}

\author{\speaker{Sean Griffin} for the AMEGO Team$^\dagger$\\
        NASA GSFC / UMD\\
        E-mail: \email{sean.griffin@nasa.gov}\\
        \llap{$^\dagger$}\href{https://asd.gsfc.nasa.gov/amego/}{https://asd.gsfc.nasa.gov/amego/}
        }
        
\abstract{
The gamma-ray sky from several hundred \SI{}{\keV} to $\sim$ a hundred \SI{}{\MeV} has remained largely unexplored due to the challenging nature of detecting gamma rays in this regime. 
At lower energies, Compton scattering is the dominant interaction process whereas at higher energies pair production dominates, with a crossover at about \SI{10}{\MeV} depending on the material. 
Thus, an instrument designed to work in this energy range must be optimized for both Compton and pair-production events. 
The All-sky Medium Energy Gamma-ray Observatory (AMEGO) is a NASA Probe-class mission concept being submitted to the Astro2020 review.
The instrument is designed to operate from \SI{200}{\keV} to $>\SI{10}{\GeV}$ and is made of four major subsystems: a plastic anti-coincidence detector for rejecting cosmic-ray events, a silicon tracker for tracking pair-production products and tracking and measuring the energies of Compton-scattered electrons, a CZT calorimeter for measuring the energy and location of Compton scattered photons, and a CsI calorimeter for measuring the energy of the pair-production products at high energies. 
A prototype instrument comprising each subsystem is currently being developed in preparation for a beam test and a balloon flight. 
In this contribution we discuss the current status of the prototype subsystems.
}

\FullConference{36th International Cosmic Ray Conference -ICRC2019-\\
		July 24th - August 1st, 2019\\
		Madison, WI, U.S.A.}

\usepackage{cleveref}
\begin{document}
\section{Introduction}

To date, the region of the electromagnetic spectrum \SIrange[scientific-notation = engineering,exponent-to-prefix = true, round-mode=figures]{1e5}{100e6}{\eV} has had very limited observations. 
This so-called ``impossible'' energy range is particularly difficult due to the fact that there are two competing photon interactions (Compton scattering and pair production) with interaction cross sections that crossover around  $\SI{10}{\MeV}$.


The \emph{All-sky Medium Energy Gamma-ray Observatory} (AMEGO) \cite{AMEGOWhitePaper} is a Probe-class ($\$\SI{1}{B}$) NASA mission concept being proposed to the Astro2020 Decadal Survay. 
The instrument is designed to operate from $\sim\SI{200}{\keV}$ to $>\SI{10}{\GeV}$ and is optimized for continuum flux sensitivity across a broad energy range with a wide field of view. 
To cover this broad energy range, \amego\ is designed to be sensitive to both Compton and pair-production interactions.
The instrument has four subsystems: a silicon microstrip tracker for measuring particle tracks and the energy of Compton scattered electrons, CZT imaging calorimeter for measuring the energy and position of Compton scattered photons, a CsI(Tl) log calorimeter for measuring the energy of $e^-/e^+$ pairs for higher energy interactions, and a plastic scintillator anticoincidence detector as the first line of defense against cosmic ray interactions. 

\amego\ is focused on the upcoming field of multimessenger astronomy. 
The recently discovered electromagnetic counterpart to the binary neutron star merger event GW170817A / GRB 170817A \cite{2017ApJ...848L..12A} and the detection of neutrinos from TXS~0506+056 \cite{2018Sci...361.1378I} during a gamma-ray flare are prime examples of where gamma-ray observations were critical to understanding the underlying physics of these highly energetic sources. 
In addition, \amego's CZT calorimeter enables nuclear line spectroscopy which allows for direct probes into element formation \cite{2019BAAS...51c...2T} and studies of the galactic \SI{511}{\keV} emission \cite{2019BAAS...51c.256K}.


In preparation for \amego, we are developing prototype versions of each detector subsystem. 
This instrument, known as \compair\ (Compton/Pair-production), will function as a proof-of-principle for \amego.
This work is supported by NASA APRA funding with grants supporting the development of the silicon tracker, beam, and balloon tests (PI: Julie McEnery), CZT calorimeter (PI: D. Thompson), and CsI calorimeter (PI: J. E. Grove). 
Descriptions and progress on the development of prototypes of each subsystem are given in the following sections. 

\section{Instrumentation}

The \amego\ design philosophy is to use many modular, highly segmented components. 
This both allows for multiple, parallel assembly lines for various components and the easy production of spares, and enables high-precision measurements of interactions within the detectors. 

\subsection{Double-sided Silicon Strip Tracker}

The silicon tracker is the first point of interaction for an incident gamma ray; the silicon acts as both the scattering target for Compton events and the conversion material for pair production. 
Unlike traditional pair-production telescopes (\eg\ \fermiLAT, \emph{AGILE}) \amego\ cannot use a passive material such as tungsten foil to promote interactions since the passive material would severely hinder Compton reconstruction. 

A Compton telescope reconstructs the arrival direction of photons using the interaction position and energy of both the scattered electron and photon. 
This allows the arrival direction of the photon to be reconstructed as a ring on the sky, requiring multiple overlapping rings to identify sources. 
However, if the scatter \emph{direction} of the electron is also measured (a so-called ``tracked event''), it is possible to use the kinematics of the interaction to reduce this direction uncertainty to an arc on the sky. 

Compton scattered electrons are typically low energy and quickly deposit their energy in the particle tracker. 
As such, double-sided silicon detectors (DSSDs) are required so that both the $x$- and $y$-coordinate of the interaction are measured simultaneously. 
In addition, analogue readout of the silicon is required in order to measure the energy of the Compton-scattered electron. 
In the lowest part of the the pair regime ($E \geq \SI{15}{\MeV}$), the energy deposition in the silicon can be non-negligible, so the analogue readout will also help with the reconstruction of low-energy pair events.

Each \compair\ DSSD is \SI{10x10}{\cm} and \SI{500}{\micro\meter} thick.
The thickness is a careful optimization between providing enough material to promote pair-production and being thin enough to reduce the chances of multiple scatterings within a detector layer. 
Each detector side has 192 orthogonally oriented strips on a \SI{510}{\micro\meter} pitch. 

\begin{figure}[hbtp]
    \centering
    \includegraphics[clip, trim={7cm 0 0 5cm}, width=0.75\textwidth]{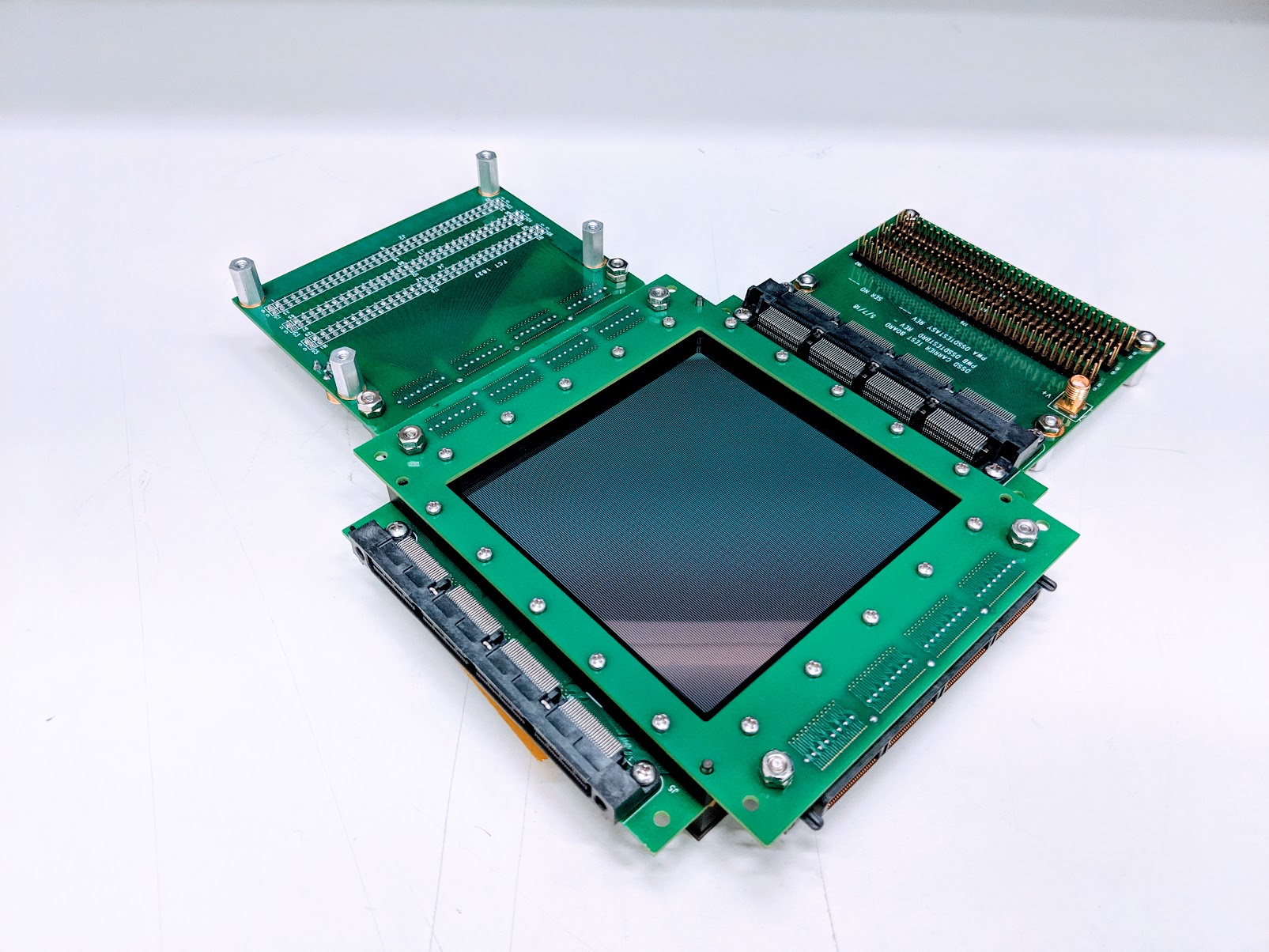}
    \caption{A \compair\ DSSD in a test carrier with breakout boards connected to each side.}
    \label{fig:DSSD}
\end{figure}

A custom detector carrier, shown in \Cref{fig:DSSD}, has been developed to interface with the detectors without the need for wire-bonding. 
This carrier allows for different front-end electronics to be easily interfaced with the silicon which simplifies testing using multiple readout systems such as bench-top electronics for cross-calibration.
Furthermore, the design allows for multiple detectors to be daisy-chained, approximating a detector ``ladder'' which is closer to the full \amego\ design. 
The noise in a detector ladder scales with the number of detectors in it, so a good understanding of the performance of a ladder is crucial to \amego.
A more traditional wire-bonded carrier has also been designed in order to understand any noise contributions due to the carriers. 

The DSSDs are read out using custom front-end electronics (FEE) based on the IDEAS VATA460.3 ASIC; the device is in the process of being qualified \cite{2019arXiv190209380G}. 
The ASIC is a combined charge-sensitive preamplifier combined with a 10-bit ADC and supports both positive and negative signals. 
Each ASIC has 32 channels, thus, each FEE board will have six ASICs. 
A tracker layer requires two FEE boards (one for each side of the detector) and a digital back-end board containing an FPGA which handles communication with the instrument trigger module, data acquisition from the ASICs, telemetry, and offloading data to the main data acquisition computer. 

\subsection{CZT Imaging Calorimeter}

\begin{figure}
    \centering
    \includegraphics[clip, trim={2cm 13cm 0 2.7cm}, width=0.75\textwidth]{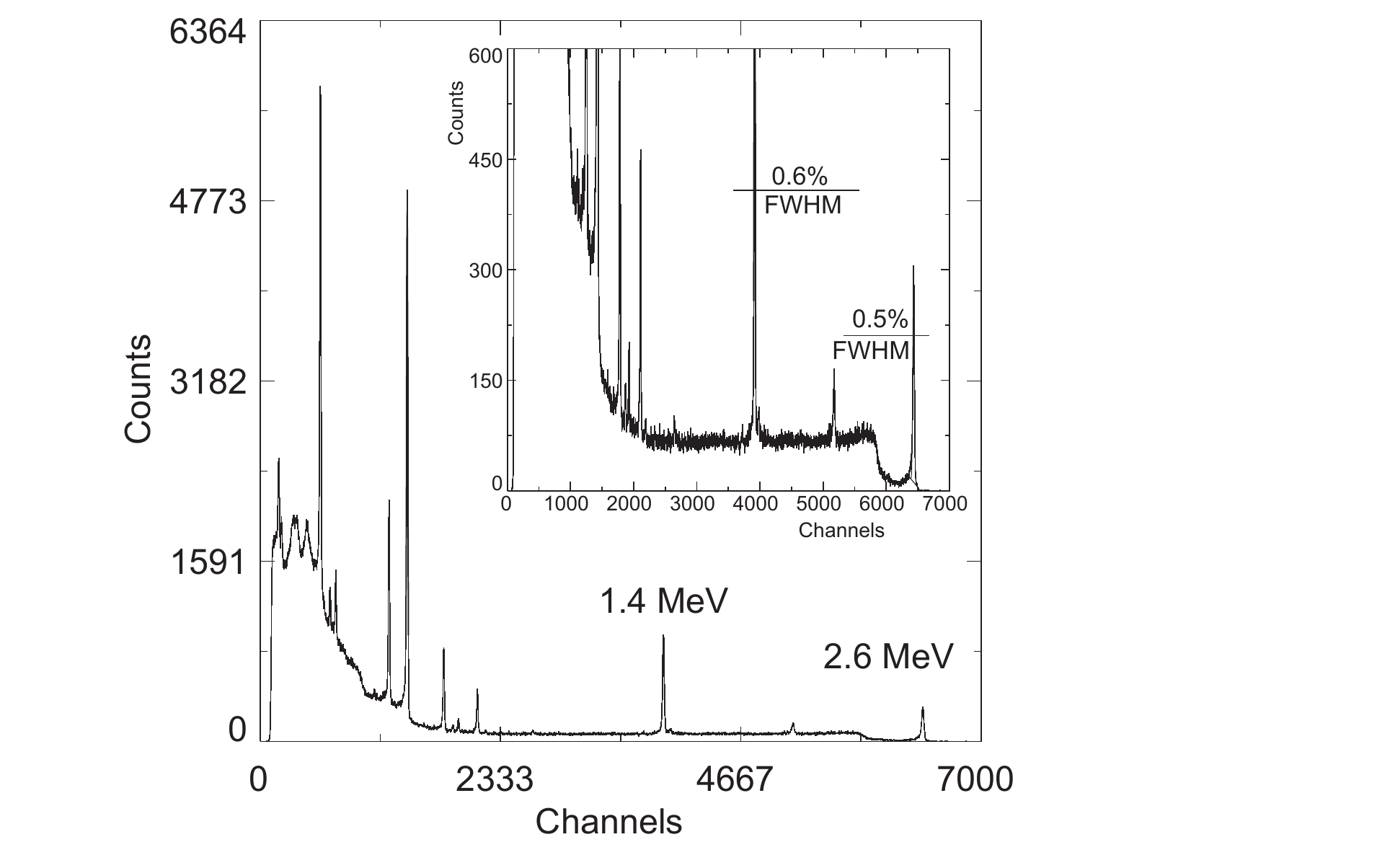}
    \caption{Energy spectrum measured by \SI{8x8x32}{\mm} CZT bar exposed to \ce{^{232} U}.}
    \label{fig:CZT_eres}
\end{figure}

Cadmium-Zinc-Telluride is a room-temperature semiconductor detector material with exquisite position and energy resolution.
It is crucial to maximizing performance in the Compton regime where event reconstruction relies heavily on a precise measurement of the Compton scattered photon's energy and scatter direction. 
Furthermore, the high energy resolution ($\Delta E = 0.5\%$ FWHM at \SI{662}{\keV}, which approaches the stochastic $e^-/h^+$ production limit) enables gamma-ray spectroscopy. 
The \amego\ CZT imaging calorimeter is based on arrays of position-sensitive virtual Frisch-grid detectors recently introduced by Brookhaven National Laboratory \citep{2014ITNS...61.2567L}, example performance of the bars can be seen in \Cref{fig:CZT_eres}. 

The base element in the CZT calorimeter is a \SI{0.6 x 0.6 x 3}{\cm} CZT bar.
Metal pads are attached to the shell near the bar's anodes; the relative amplitudes of the signals read out from the pads are used to evaluate the $x/y$ coordinate of interactions within the detector, and the drift time and cathode-to-anode signal ratio are used to independently evaluate the $z-$ coordinate of the interaction location. 

The bars are arranged in a $4 \times 4$ module and connected to a readout ASIC (the AVG2 \cite{2019NIMPA.940....1V}) and FPGA module which controls readout. 
An individual module is \SI{2.5x2.5}{\cm}, the \compair\ prototype will use an array of $4\times 4$ modules, matching the footprint of a DSSD.
This is in keeping with the concept of modularity for the \amego\ design. 
In May 2018, the first prototype CZT module was tested in a gamma-ray beam at the High Intensity Gamma-ray Source (HIGS) facility at Duke University\footnote{\href{http://www.tunl.duke.edu/web.tunl.2011a.higs.php}{http://www.tunl.duke.edu/web.tunl.2011a.higs.php}}. 
The calorimeter design is currently being refined; assembly has begun on the next modules and the associated front-end electronics. 

\begin{figure}[hbtp]
    \centering
    \includegraphics[height=0.35\textwidth]{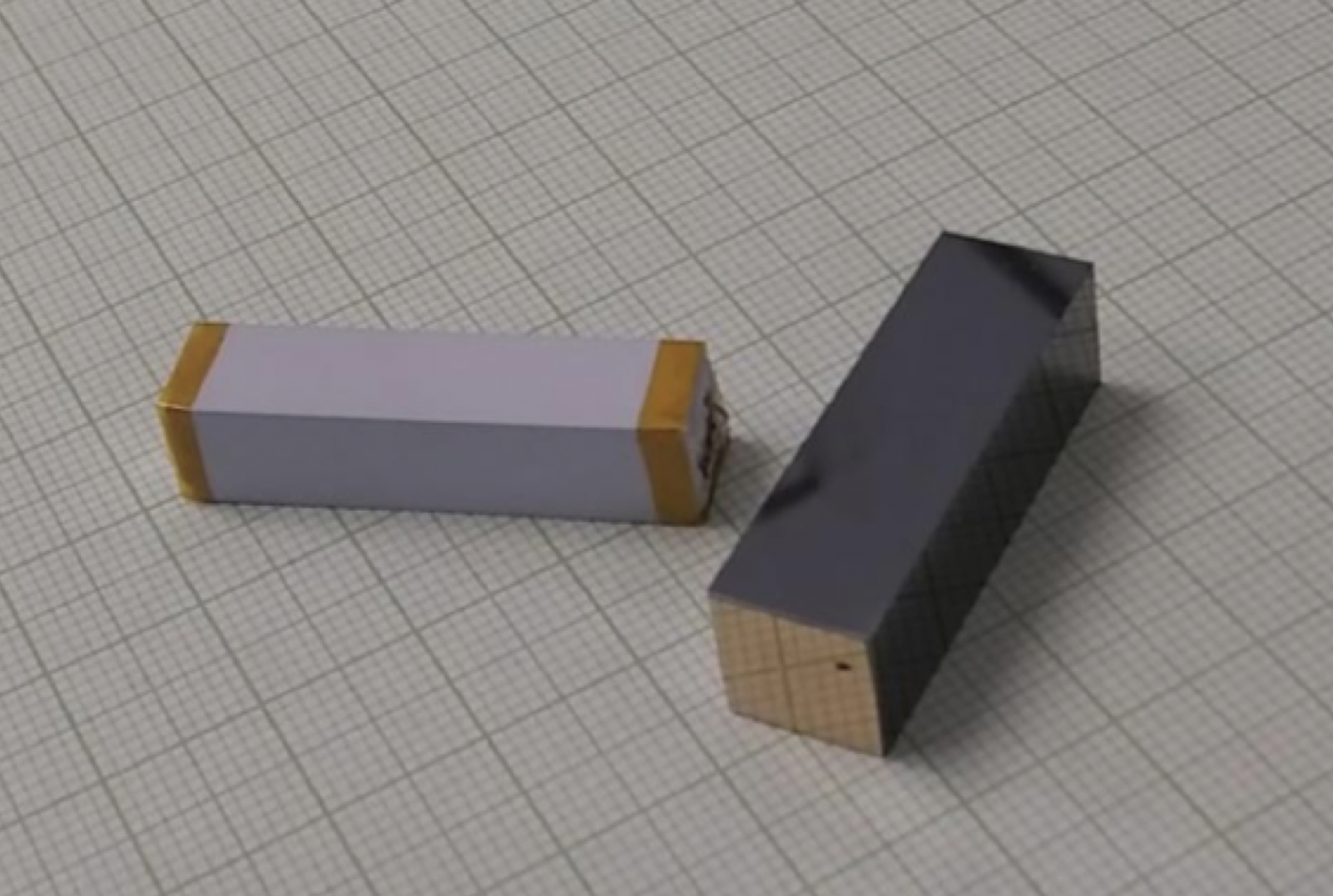}
    \includegraphics[height=0.35\textwidth]{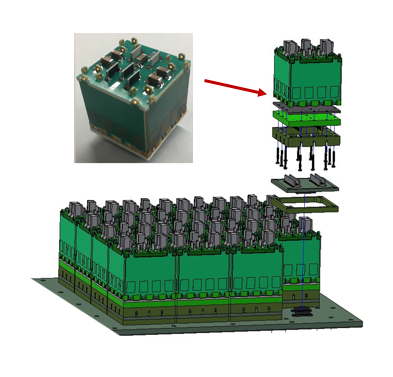}
    \caption{\textbf{Left}: Photo of a bare and wrapped CZT crystal. 
    \textbf{Right}: A schematic view of the assembly of a CZT module.}
    \label{fig:CZT_module_assembly}
\end{figure}

\subsection{CsI Calorimeter}

The \compair\ CsI calorimeter (and by extension \amego) leverages the experience gleaned from the \fermiLAT. 
The CsI calorimeter is made of four layers of CsI(Tl) bars arranged in a hodoscopic configuration. 
Each bar is read out using silicon photomultipliers (SiPMs); the light asymmetry between signals on the two ends provide a 1-D interaction location; the other two dimensions are based on the geometry of the instrument.

SiPMs are ideal for space applications due to their compact size (typical arrays can be of order \SI{1}{\cm\squared} and only $\sim$ a few \SI{}{\mm} thick) and low operating voltage (\SIrange{30}{60}{V} typical), while still having gains comparable to classical photomultiplier tubes ($10^5 - 10^6$). 
The \compair\ SiPMs are read our using the ROSSPAD data acquisition system which comprises four IDEAS SIPHRA ASICs (IDE3380) \cite{siphra} and a dedicated FPGA.

\begin{figure}[hbtp]
    \centering
    \includegraphics[height=0.375\textwidth]{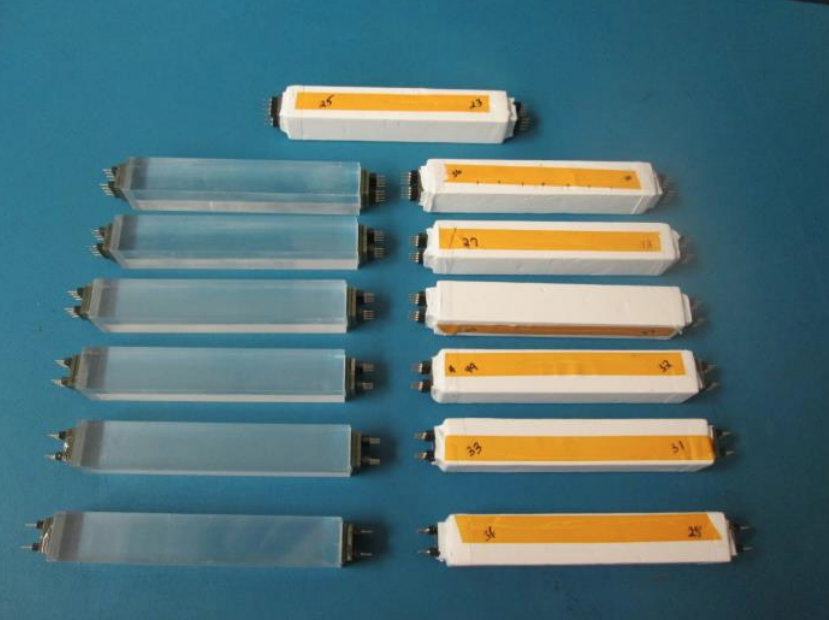}
    \includegraphics[height=0.375\textwidth]{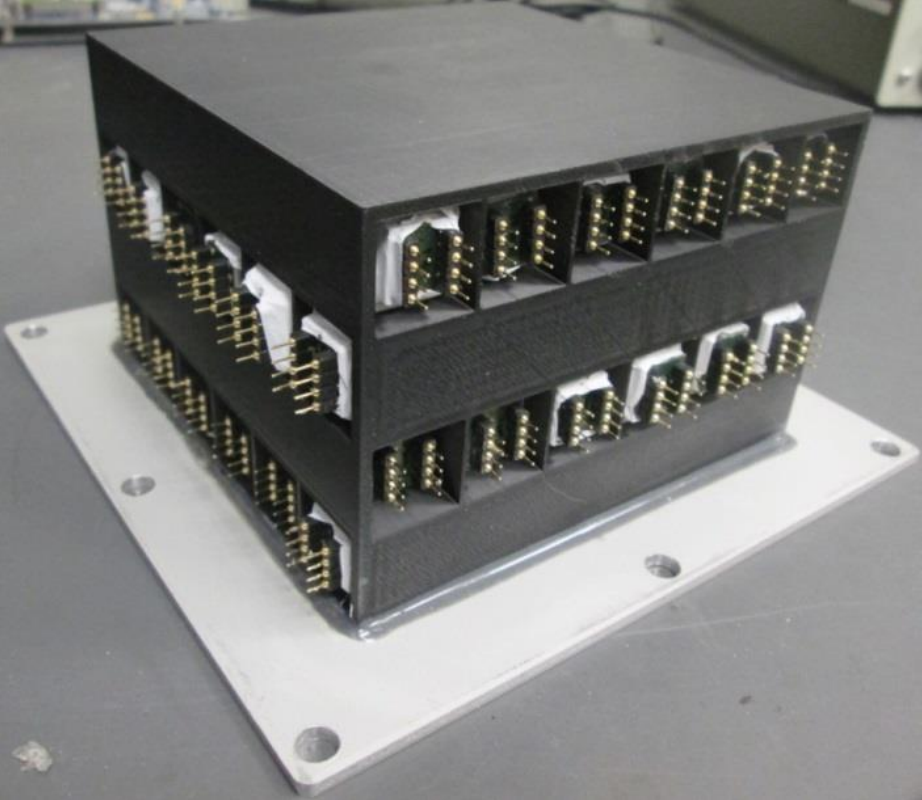}
    \caption{\textbf{Left:} CsI crystal bars with SiPMs bonded to either side. 
    The bars on the right have been wrapped in Tetratek to ensure good light collection.   
    \textbf{Right:} The prototype 24-element CsI hodoscopic calorimeter using SiPM readout. 
    }
    \label{fig:compair_cal}
\end{figure}

The prototype dimensions have been chosen to be commensurate with the dimensions of a single-detector-per-layer prototype tracker and a \SI{10 x 10}{\cm} CZT calorimeter. 
Each of the four hodoscope layers has six \SI{16.7 x 16.7 x 100}{\mm} scintillator bars resulting in a thickness of $\sim 3.7 X_0$.
The surface of each bar is roughened to increase the light asymmetry of both sides; this greatly improves position resolution at a minimal cost of total light yield. 
The bar is then wrapped in a diffuse reflective material (Tetratex) in order to maximise light collection and \SI{6x6}{\mm} SiPM arrays are bonded to the end of each scintillator bar. 
A photo of the prototype hodoscope is given in \Cref{fig:compair_cal} and more details can be found in \cite{2019arXiv190105828W}.
For full \amego, the CsI bars will be closer to $\sim \SI{40}{\cm}$, commensurate with the footprint of a full tower.

\subsection{Anticoincidence Detector}

The anticoincidence detector (ACD) is critical for rejecting events corresponding to cosmic-ray interactions in the instrument which dominate over gamma-ray interactions by up to five orders of magnitude. 
The baseline \amego\ ACD is five panels covering four sides and the top of the instrument; the \compair\ design is similar. 
Like the \ce{CsI} calorimeter, the ACD is read out using SiPMs and a SIPHRA ASIC. 
At high energies, interactions in the CZT and CsI calorimeters can backscatter and produce falase vetos in the ACD; these can be identified by segmenting the ACD scintillators \cite{2007APh....27..339M}; this requirement is currently under investigation.

\subsection{Trigger and Data Acquisition}

\compair\ uses a trigger module to search for coincident ``hit'' signals in each subsystem. 
A typical trigger will correspond to two hits in the tracker (corresponding to an $x-$ and $y-$ coordinate) and a hit in either of the calorimeters. 
Due to the high position sensitivity of the CZT calorimeter, it is also possible to reconstruct Compton events which occur only in the calorimeters. 
\compair\ will employ several trigger engines which will allow for pre-scalings to be applied to each trigger type such that the data stream will not be dominated by any given trigger type and allow for a more even sampling of the different event types. 

\section{Future Work and Conclusions}

\amego\ is a powerful \SI{}{\MeV} gamma-ray telescope that will open a window on the universe with unprecedented sensitivity. 
The prototype, \compair, is under development. 
Each subsytem is in the process of being assembled and integration will begin in the end of 2019. 
Once integrated, \compair\ will be beam tested at HIGS. 
HIGS can produce gamma rays from \SIrange{2}{100}{\MeV}; as such it is possible to test \compair\ in both the Compton and pair-production regimes. 
Furthermore, the beam supports circular and linear polarization which could provide proof-of-principle polarization measurements. 
We are currently targeting a short-duration balloon flight for \compair\ in 2021.

\bibliography{bib}

\providecommand{\href}[2]{#2}\begingroup\raggedright\begin{thebibliography}{10}

\bibitem{AMEGOWhitePaper}
J.~{McEnery}, J.~{Abel Barrio}, I.~{Agudo}, M.~{Ajello}, J.-M. {{\'A}lvarez},
  S.~{Ansoldi} et~al., \emph{{All-sky Medium Energy Gamma-ray Observatory:
  Exploring the Extreme Multimessenger Universe}}, {\emph{Astro2020 APC White
  Paper} (2019) } [\href{https://arxiv.org/abs/1907.07558}{{\ttfamily
  1907.07558}}].

\bibitem{2017ApJ...848L..12A}
B.~P. {Abbott}, R.~{Abbott}, T.~D. {Abbott}, F.~{Acernese}, K.~{Ackley},
  C.~{Adams} et~al., \emph{{Multi-messenger Observations of a Binary Neutron
  Star Merger}}, \href{https://doi.org/10.3847/2041-8213/aa91c9}{\emph{ApJL}
  {\bfseries 848} (2017) L12}
  [\href{https://arxiv.org/abs/1710.05833}{{\ttfamily 1710.05833}}].

\bibitem{2018Sci...361.1378I}
{IceCube Collaboration}, M.~G. {Aartsen}, M.~{Ackermann}, J.~{Adams}, J.~A.
  {Aguilar}, M.~{Ahlers} et~al., \emph{{Multimessenger observations of a
  flaring blazar coincident with high-energy neutrino IceCube-170922A}},
  \href{https://doi.org/10.1126/science.aat1378}{\emph{Science} {\bfseries 361}
  (2018) eaat1378} [\href{https://arxiv.org/abs/1807.08816}{{\ttfamily
  1807.08816}}].

\bibitem{2019BAAS...51c...2T}
F.~{Timmes}, C.~{Fryer}, F.~{Timmes}, A.~L. {Hungerford}, A.~{Couture},
  F.~{Adams} et~al., \emph{{Catching Element Formation In The Act ; The Case
  for a New MeV Gamma-Ray Mission: Radionuclide Astronomy in the 2020s}},  in
  \emph{Bulletin of the American Astronomical Society}, vol.~51, p.~2, May,
  2019, \href{https://arxiv.org/abs/1902.02915}{{\ttfamily 1902.02915}}.

\bibitem{2019BAAS...51c.256K}
C.~{Kierans}, J.~F. {Beacom}, S.~{Boggs}, M.~{Buckley}, R.~{Caputo},
  R.~{Crocker} et~al., \emph{{Positron Annihilation in the Galaxy}},  in
  \emph{Bulletin of the American Astronomical Society}, vol.~51, p.~256, May,
  2019, \href{https://arxiv.org/abs/1903.05569}{{\ttfamily 1903.05569}}.

\bibitem{2019arXiv190209380G}
S.~{Griffin} and {the AMEGO Team}, \emph{{Development of a Silicon Tracker for
  the All-sky Medium Energy Gamma-ray Observatory Prototype}},  in \emph{IEEE
  NSS}, Feb, 2019, \href{https://arxiv.org/abs/1902.09380}{{\ttfamily
  1902.09380}}.

\bibitem{2014ITNS...61.2567L}
K.~{Lee}, A.~{Bolotnikov}, S.~{Bae}, U.~{Roy}, G.~{Camarda}, M.~{Petryk}
  et~al., \emph{{New Virtual Frisch-Grid CdZnTe Detector Design With
  Sub-Millimeter Spatial Resolution}},
  \href{https://doi.org/10.1109/TNS.2014.2348572}{\emph{IEEE Transactions on
  Nuclear Science} {\bfseries 61} (2014) 2567}.

\bibitem{2019NIMPA.940....1V}
E.~{Vernon}, G.~{De Geronimo}, A.~{Bolotnikov}, M.~{Stanacevic}, J.~{Fried},
  L.~O. {Giraldo} et~al., \emph{{Front-end ASIC for spectroscopic readout of
  virtual Frisch-grid CZT bar sensors}},
  \href{https://doi.org/10.1016/j.nima.2019.05.047}{\emph{Nucl. Instrum.
  Methods Phys. Res. A} {\bfseries 940} (2019) 1}
  [\href{https://arxiv.org/abs/1904.01529}{{\ttfamily 1904.01529}}].

\bibitem{siphra}
D.~Meier, J.~Ackermann, A.~Olsen, H.~Kristian, H.~Berge, A.~Hasanbegovic
  et~al., \emph{{SIPHRA} 16-channel silicon photomultiplier readout {ASIC}},
  in \emph{AMICSA and DSP}, June, 2016,
  \href{https://doi.org/10.13140/RG.2.1.1460.8882}{DOI}.

\bibitem{2019arXiv190105828W}
R.~S. {Woolf}, J.~E. {Grove}, B.~F. {Phlips} and E.~A. {Wulf},
  \emph{{Development of a CsI:Tl calorimeter subsystem for the All-Sky
  Medium-Energy Gamma-Ray Observatory (AMEGO)}},  in \emph{IEEE NSS}, Jan,
  2019, \href{https://arxiv.org/abs/1901.05828}{{\ttfamily 1901.05828}}.

\bibitem{2007APh....27..339M}
A.~A. {Moiseev}, R.~C. {Hartman}, J.~F. {Ormes}, D.~J. {Thompson}, M.~J.
  {Amato}, T.~E. {Johnson} et~al., \emph{{The anti-coincidence detector for the
  GLAST large area telescope}},
  \href{https://doi.org/10.1016/j.astropartphys.2006.12.003}{\emph{Astroparticle
  Physics} {\bfseries 27} (2007) 339}
  [\href{https://arxiv.org/abs/astro-ph/0702581}{{\ttfamily
  astro-ph/0702581}}].

\end{thebibliography}\endgroup


\section*{Acknowledgements}

We are grateful to the accelerator physics group led by Prof. Y.K. Wu at HIGS, Triangle Universities Nuclear Laboratory for developing and producing an excellent gamma-ray beam for the tests. The HIGS facility is supported in part by U.S. Department of Energy grant DE-FG02-97ER41033.

\end{document}